# Prediction of optimal catalysts for a given chemical reaction

Hervé Toulhoat,*[a] Pascal Raybaud [b]

[a] Sorbonne Université, UPMC, CNRS, Laboratoire de Réactivité de Surface, 4 Place Jussieu, 75005, Paris, France.

[b] IFP Energies nouvelles, Rond-point de l'Echangeur de Solaize, BP 3, 69360, Solaize, France.

* Corresponding author.

**Abstract**

We reveal a correlation between the M-X bond energy descriptor EMX for the optimal catalyst in a family of stoichiometry MiXj, and an intensive quantity defined as the standard enthalpy of the catalyzed reaction normalized to one mole of element X transferred by this reaction from reactants to products. M is a transition element, and the stoichiometry MiXj is fixed at the solid/fluid interface by the reaction conditions. We illustrate this for a relevant set of reactions involved in solar energy and industrial applications such as oxygen evolution, oxygen reduction and hydrogen evolution in electrolysis, hydrodesulfurization of thiophene, methanation of CO, hydrogenations of aromatics and alkenes, selective oxidation of methane, and ammonia synthesis and decomposition. We propose a quantitative model to explain this unexpected connection: this key finding and its interpretation should accelerate in silico discovery of catalysts.

## 1. Introduction

Although most industrially important catalysts known so far were serendipitous discoveries further optimized by trial and errors, achieving rational design of catalysts has been one of the major challenges in chemistry for more than a century. Besides, the current global context is characterized by an ever-growing demand in energy and raw materials, driven by population growth and legitimate expectations of economic development. Simultaneously, there are rising concerns about global warming caused by greenhouse gases emission and environmental impact of manufactured chemicals. These combined incentives call for focused efforts to discover efficient new catalysts based on non-ecotoxic and earth-abundant elements. Catalytic processes are indeed recognized as keys to a sustainable future, through the production of “solar fuels” and bio-based chemicals.

The work by Kamil Klier[1] was among the early attempts to describe periodic trends in catalytic activity patterns by a quantity assimilable to a relevant surface bond strength. This author proposed to define the binding energy of oxygen to a transition metal oxide surface by the standard enthalpy of the reaction 1/n $Me_xO_y$ + ½ $O_2$ = 1/n $Me_xO_{y+n}$ , while showing the oxidation-reduction potential of solid oxides at a temperature T may be defined as the free energy change of this reaction at this temperature $-\Delta F_T^0$ easily deduced from tables of experimental thermodynamic dat. He showed that specific activities of transition metal oxides in hydrogen oxidation at 300 °C reported by Boreskov and Golodets plotted against $-\Delta F_{600K}^0$ arrange themselves along a volcano curve culminating for cobalt and copper oxides.

It has been shown recently that for a given reaction, optimal catalysts can be quite generally identified thanks to so-called "volcano plots" which unequivocally relate specific activities (turn over frequencies) to "descriptors" or quantitative measures related to the affinity between reactants and catalytic surfaces[2,3]. It is indeed a manifestation of the well-known Sabatier principle[4]. Modern first-principles computational methods allied to available massive supercomputing power, are closing chemical accuracy in the simulation of the electronic and structural properties of chemical systems represented by models comprising now routinely up to a few hundreds of atoms spanning the entire periodic table[2]. Although average errors of calculated bond energies are still over 20 $kJ.mol^{-1}$, this methodology is accurate enough to identify key chemical trends and structure-activity relationships[5]. While these modern computational chemistry techniques allow in principle to establish free energy profiles along reaction complex reaction pathways at surfaces, and hence provide directly the relevant interaction energies and activation barriers, it remains an immense task even for a single activity pattern since case by case proper representations of the solid surfaces as expressed operando must be built and validated against experimental characterizations operando, while the latter are still in their infancy. By contrast any approach based on descriptors like adsorption energies on model surfaces, or characteristics of the electronic structure of the latter (e.g. density of states) implies some empirical character, only justified by its success in representing a volcano pattern and allowing predictions for solids not included in the “training set”. In a sense this class of approaches is akin to machine-learning and may be systematically improved in this spirit as recently demonstrated[6]. In particular, several examples of volcano plots have been reported for industrially and environmentally significant reactions, where bond-strength descriptors are computed upon solving the Schrödinger equation in the framework of the density functional theory (DFT). Moreover, the approach was proven predictive in the sense that "chemical interpolations", that is combining un-active or weakly active materials described to belong to the "sides" of a volcano into a mixed compound described to belong to the "summit", were verified to be highly active[2,3]. Therefore a combinatorial in silico screening of candidate materials for catalysis is affordable in principle. However, the optimal bond energy for a given reaction has been so far unpredictable by the sole remedy of theoretical approach, so that, building the volcano plot necessitates an experimental activity pattern, namely a very consistent set of turnover frequencies measured in the same conditions for a set of comparable catalysts differing as far as possible only by one chemical element involved. For instance, mono-disperse transition metal nanoparticles supported on a high surface area silica[7], or transition metal sulfides nanoparticles supported on a high surface area alumina[8].

Collecting such data thus involves a considerable amount of laboratory skills, time and cost, so that a theoretical guide to predict directly the optimal catalyst for a given reaction is highly desirable. Recently, Chen et al. proposed a method which they named “the half-principle” aiming at fulfilling this program[9]. This method relies on the identification of a “critical intermediate” I in adsorbed phase, the chemical potential of which on the optimal catalyst would be the arithmetic mean of those of lumped reactants and lumped products so that $\mu_I^{opt} = 1/2\,(\mu_R + \mu_P)$. These authors derive this relation however only for a very particular and simplified model of the catalytic process, assuming merely two consecutive activated steps, adsorption R(g)→I(ads), and desorption I(ads)→P(gas). This model cannot obviously account for the diversity of situations encountered in heterogeneous catalysis, as exemplified by many surface science and DFT studies which have located rate determining steps in adsorbed phase. In practice, the identification of the critical intermediate remains more intuitive than rational, and the further determination of $\mu_I^{opt}$ relies on scaling relationships resulting from systematic heavy DFT calculations of adsorption free energies on model surfaces of potential catalysts. This approach however points out that "the optimal catalyst is mainly determined by thermodynamic parameters", as a first approximation, and recognize in this respect as a precursor the multiplet theory of Balandin[10].

In what follows, we also discover a simple connection between a descriptor of the optimal catalyst of a given reaction and the thermodynamics of this reaction, but through a quite different approach, independent of restrictive hypotheses on the reaction pathway. Our demonstration is first supported by showing empirically a simple relationship between the bond energy describing the optimal solid catalyst, and the absolute value of the standard enthalpy for twelve catalysed reactions relevant in refining, petrochemistry and solar fuel applications. We then propose a rational interpretation of this relationship.

This discovery allows in principle to circumvent the need of reference experimental activity patterns, and therefore to greatly expand the applicability of in silico screening of catalytic formulation.

## 2. Methods

The "Yin–Yang" algorithm has been previously described[3]. In the variant used here, absolute values of total energies per unit cell are computed for:

- the optimized bulk compound $M_iX_j$, $E_{Yin-Yang}$,

- the same unit cell as for the bulk with target atom M removed (X positions are frozen), $E_{Yin}$,
- the same fixed unit cell as for the $M_iX_j$ bulk from which all atoms except the target atom M have been removed (M positions are frozen), $E_{Yang}$.

Note: in the cases of materials $M^k_{i_k}X_j$ including several transition metals $M^k$ , each of them can be separately chosen as the target atom.

With $n$ the number of nearest neighbor atoms X to target atom M per unit cell, we define the bond energy descriptor $E_{MX}$ as:

$$E_{MX} = \left[E_{Yin-Yang} - \left(E_{Yin} + E_{Yang}\right)\right]/n \quad (1)$$

$E_{MX}$ can therefore be viewed as a "rebonding" energy of X to its complement in the original bulk structure $M_iX_j$. Calculation were performed through the MedeA interface[11] Structures of the relevant hydrides, carbides, nitrides, oxides, and sulfides were recovered from the Pearson and ICSD crystallographic databases[12] through the InfomaticA module in MedeA, which was also helpful in analysing these structures for the determination of n and providing them as inputs of ab initio calculations. Total energy calculations were performed with the VASP version 5.3 software[13] within the periodic density functional theory, using projected augmented wavefunctions, the PBE functional[14] in the generalized gradient approximation, and Spin Orbit Magnetic option. We decided not to include the semi-empirical U correction which has been introduced, to our knowledge, mostly to improve the prediction of band gaps for semi-conducting materials, while GGA has been shown to predict total energies and geometries rather accurately. Since in the Yin-Yang calculations the primary concern is cohesive energies, we choose to adopt GGA with spin orbit magnetic (SOM) correction systematically as good compromise between accuracy and computational speed. In view of the uncertainties affecting the experimental turnover frequencies it is not obvious that better descriptions of the volcano plots would be obtained with other levels of approximation (including U corrections, or using hybrid functionals). Yet, it might be an interesting issue to address in the future The unit cell parameters of the bulk compounds were optimized with the automatic k-point meshes generation option and default cut off energies, in order to minimize the total energy $E_{Yin-Yang}$ under the approximations made in VASP, but these parameters were kept unchanged to compute $E_{Yin}$ and $E_{Yang}$. All series of calculations were submitted through the PREDIBOND® module[15] within MedeA, which implements the Yin-Yang algorithm.

For computing the filling of the antibonding states of $e_g$ orbital parentage in the coordination sphere of transition metal cations $M_i$, the density of states (DOS) for the optimized structure of the complete bulk $M_iX_j$ structure was calculated with the VASP software at the same level of approximation as for $E_{MX}$ as described above. In each case, the partial DOS of *d* character spherically projected onto M atoms inside Wigner Seitz spheres was integrated from energy level $E_m$ up to the Fermi level $E_F$. Energy level $E_m$ corresponds to the minimum in DOS separating occupied $t_{2g}$ and $e^*_g$ states, which is in octahedral symmetry of TM cations general clearly apparent (Fig. S1.1 in ESI). This integral provides a number of d electrons $N_{e*_g@M}$ occupying orbitals of $e_g$ parentage .

## 3. Results

In the following sections 3.1 and 3.2, and in ESI, volcano plots are presented, which we have constructed from published experimental activity data in ordinates and Yin-Yang $E_{MX}$ descriptors in abscissae. Since it was clearly beyond the scope of the present work to reproduce the experimental activity data, we did our best to select consistent sets and when possible to compare data published by different authors on similar reactions and catalysts.

### 3.1 Solar fuels applications: electro-(photo-)catalysis

Volcano plots are identified for the oxygen evolution reaction (OER) in the electrocatalytic (Figure 1a) and photo-assisted (POER) (Figure 1b) modes. For that purpose, we report in ordinates of data points the experimental activities reported by Suntivich et al. for various perovskites[16], $IrO_2$ and $RuO_2$[17] as electrocatalysts, and by Harriman et al.[18] or more recently Robinson et al.[19] for various binary transition metal oxides as photo-cocatalysts. Abscissae of data points are metal-oxygen bond energies $E_{MO}$ in the bulk oxide. The optimal $E_{MO,opt}$ corresponds to 179.6 ± 1.5 and 175.7 ± 5 kJ.mol$^{-1}$ for OER and POER respectively, therefore identical within the error margin of about 3 %.

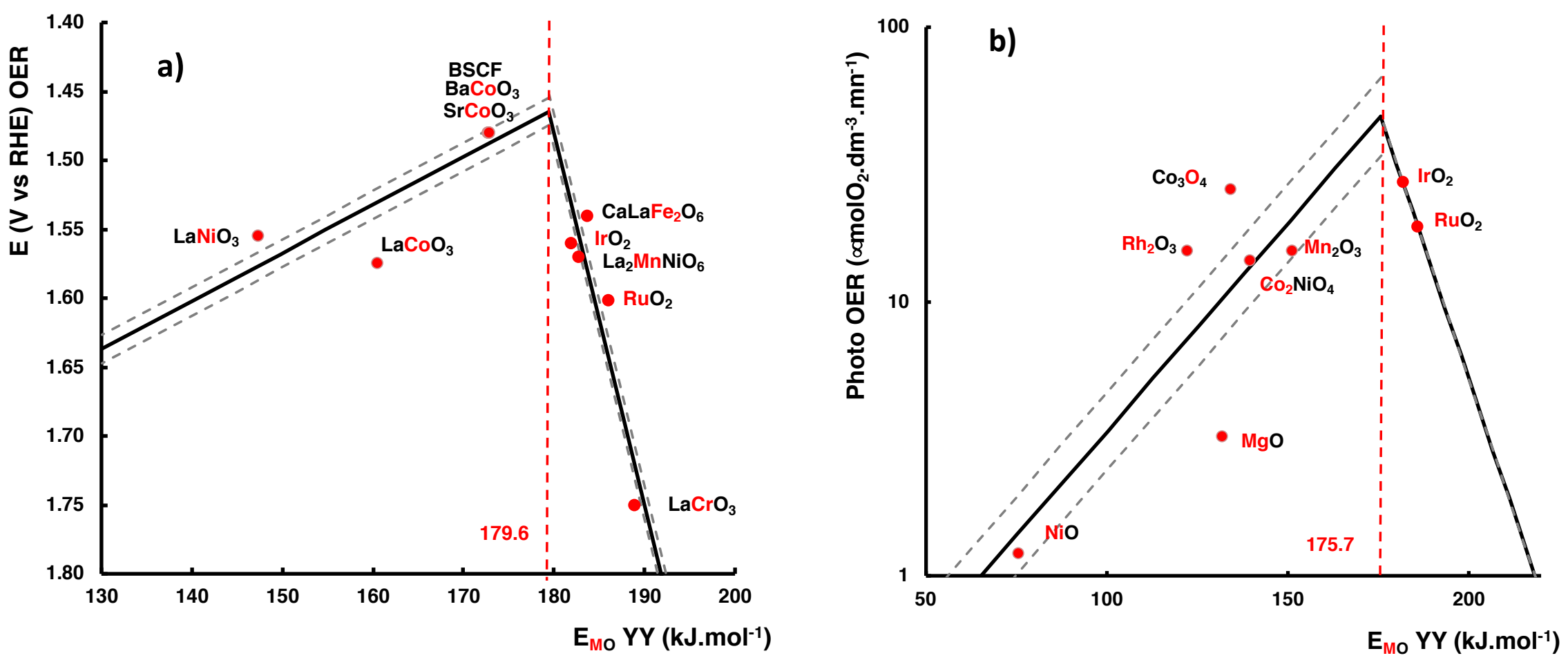

**Figure 1** Volcano plots for the electrocatalytic oxygen evolution reaction (OER) (a) and the photocatalytic OER (b). Experimental activity patterns in ordinates from [16,17] in OER, and [18, 19] in photocatalytic OER. In abscissae, DFT calculated $E_{MO}$descriptors computed according to the Yin Yang method[3] with target atoms indicated in red when necessary. The regression lines, connecting data points, materialize the left- and right- hand sides of the volcanoes. Projecting their crossing point on the X axis precisely locates the optimal values $E_{MO,opt}$. Dotted lines bracket the left- and right- hand sides regression lines by ± the standard deviation of experimental ordinates with respect to their projection on the corresponding regression line, allowing estimations of error (about 3%) bars on the coordinates of the volcano summit (see ESI section 2).

The volcano plot proposed by Suntivich et al. in [16] is described by a semi-empirical descriptor, the filling of metal antibonding states of $e_g$ orbital parentage of surface transition metal cations, which was shown by Nørskov et al.[20] to correlate with another DFT evaluation of surface metal-oxygen bond strengths. For a subset of transition metal oxides considered in this work, we revisit this electronic analysis by computing $N_{e*_g@M}$ the filling of the antibonding states of $e_g$ orbital parentage in the coordination sphere of target transition metal cations M, according to the density of states (DOS) for the optimized structure of the complete bulk $M_iX_j$ structure (see Methods). Table SI.1.1 in ESI presents the results of these calculations, while Figure 2 highlights the remarkable linear anticorrelation ($R^2$ = 0.958) obtained between descriptors $E_{MO}$ and $N_{e*_g@M}$. The intercept $N_{e*_g@M} = 10$ corresponds formally to $E_{MO} = 0$. In other terms, we recover the expected result that the bulk strength of the M-O bond is null when all available *d* electrons occupy antibonding *d* states of $e_g$ parentage.

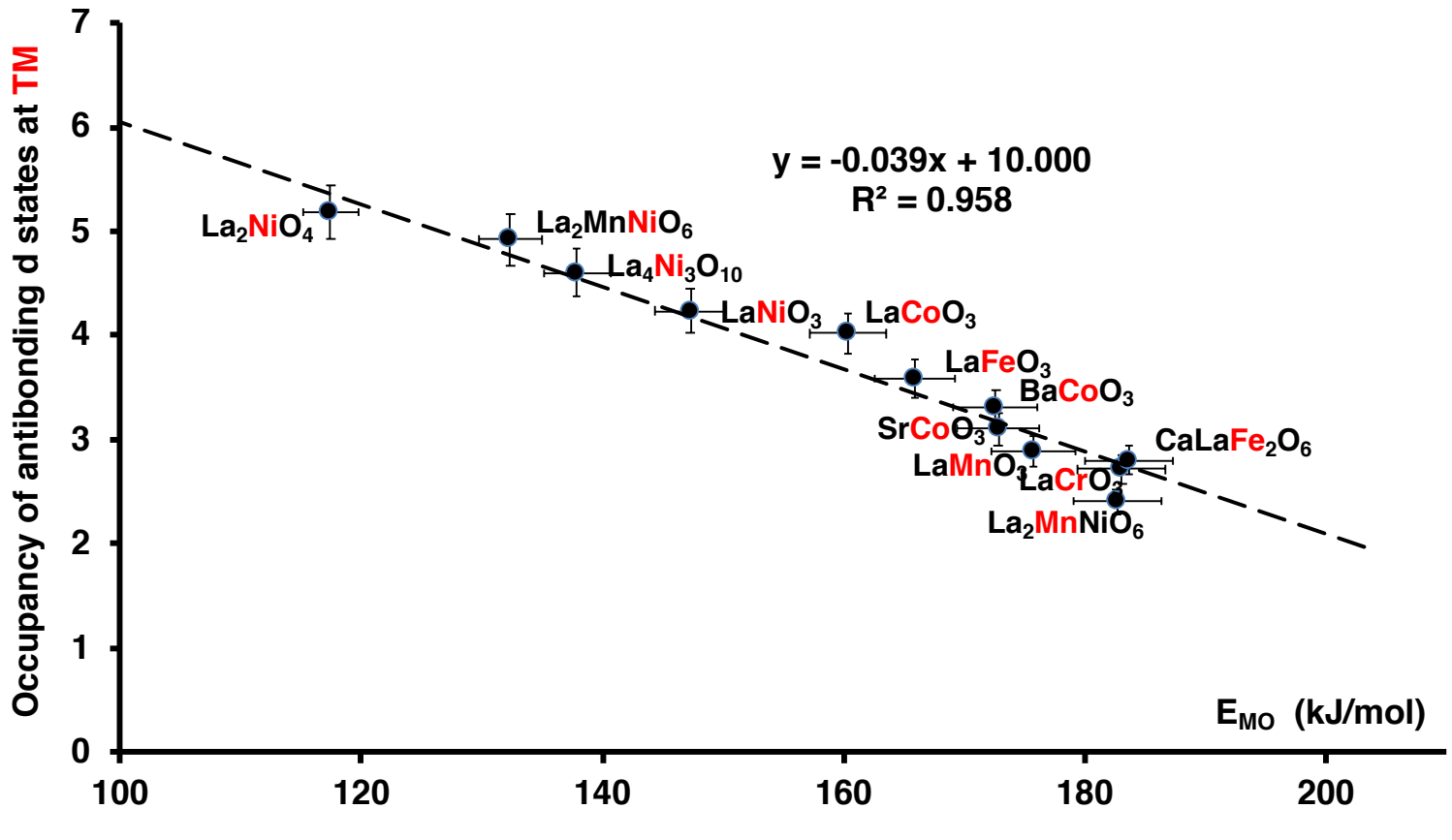

**Figure 2** Correlation between the number of d electrons $N_{e*_g@M}$ occupying orbitals of e*g parentage and $E_{MO}$ for the set of oxides presented in Table S.1.1. Errors bars correspond to 2% estimated relative errors. Dotted grey line: regression line (equation and coefficient of correlation in inset).

Moreover, Figure 3 presents the correlation of our computed descriptor $N_{e*_g@M}$ with the estimated $e_g$ occupancy assigned by Suntivich et al. in [16]. This correlation could be acceptable, if not hampered by high occupancies assigned to Fe perovskites, and the undiscriminating assignation of $e_g$ occupancy 1 to $LaMnO_3$, $LaCoO_3$ and $LaNiO_3$ while chemical intuition leads us to expect, and DFT calculations show, increasing occupancies of antibonding $e_g$ orbitals with increasing d-band filling from Mn to Ni.

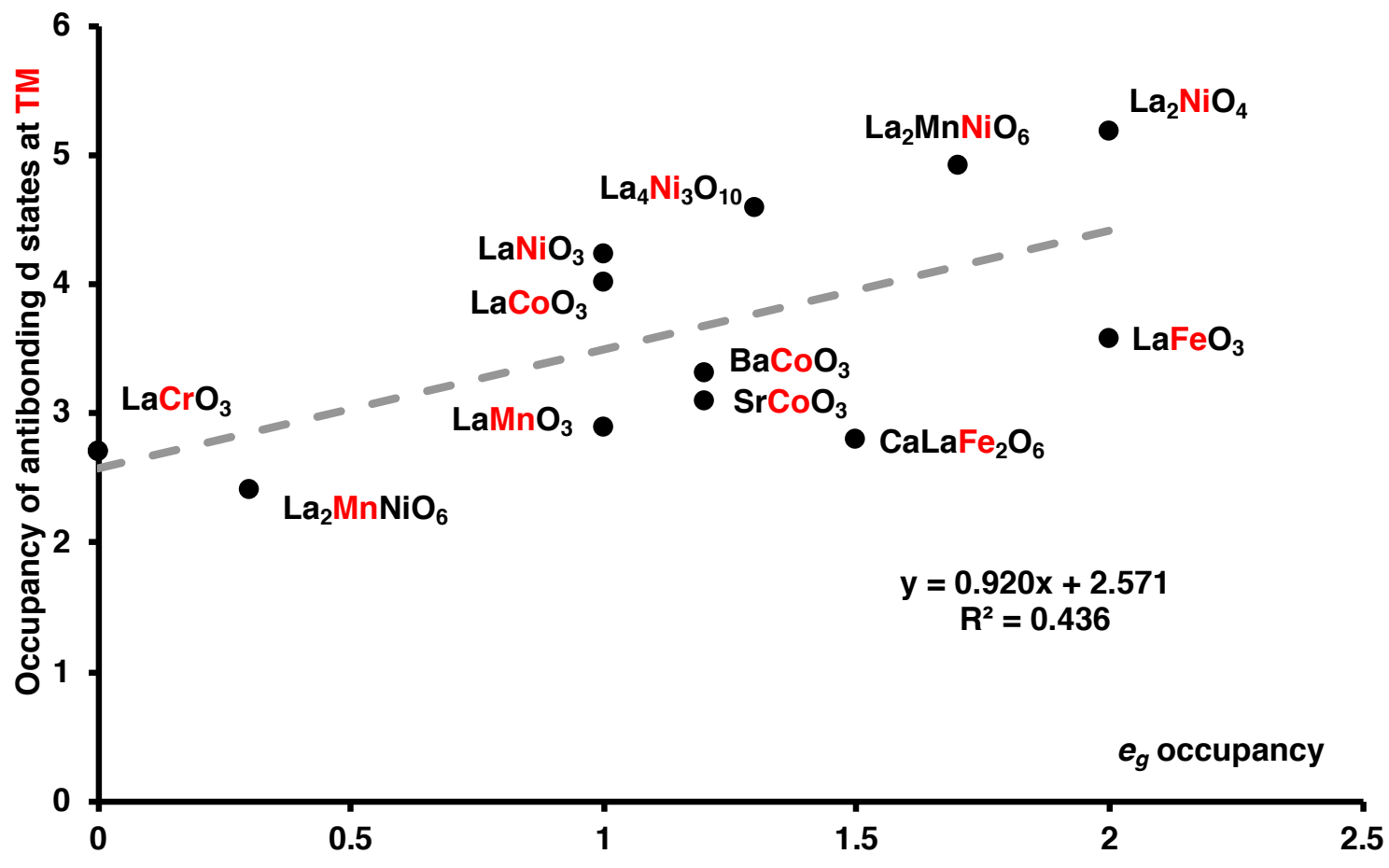

**Figure 3** Correlation between the number of d electrons $N_{e*_g@M}$ occupying orbitals of $e*_g$ parentage and $e_g$ occupancies attributed by Suntivich et al.[16] to various perovskite structures.

Thus while our results clearly support the concept introduced by Suntivich et al. that metal-oxygen bond strengths in transition metal oxides are directly related to the filling of antibonding d states of $e_g$ parentage, the DFT based Yin-Yang $E_{MO}$ appears as a more direct descriptor than the $e_g$ occupancies.

In the case of the most active material reported by Suntivich et al., a complex non-stoichiometric perovskite ($Ba_{0.5}Sr_{0.5}Co_{0.8}Fe_{0.2}O_{3-\delta}$ abbreviated BSCF), we have computed $E_{MO}$ with Co as the target atom in a large P1 unit cell affording 100 inequivalent Co ions and nominal stoichiometry, as well as in stoichiometric simple perovskites $BaCoO_3$ and $SrCoO_3$. The results are almost undistinguishable, suggesting the latter as good candidates also for OER electrocatalysis.

Similarly, we also identify volcano plots for the oxygen reduction reaction (ORR) in the electrocatalytic mode (ESI section 2). Ordinates of data points are experimental activities reported by Suntivich et al. for various perovskites, $IrO_2$ and $RuO_2$[17] (Figure

S.2.1), and by Sunarso et al.[21] for more perovskites (Figure S.2.2). The optimal metal-oxygen bond energies ($E_{MO,opt}$ = 166.9 ± 4 and 171.7 ± 4 kJ.mol$^{-1}$) are located according to the same approach as described above.

Finally, regarding the electrocatalytic hydrogen evolution reaction[22,23], a similar volcano plot is also found (ESI section 3) where the metal-hydrogen bond energy $E_{MH,opt}$ calculated in hydride materials is identified as the key descriptor. In this case, the optimal value of the descriptor is calculated at 32.9 ± 0.3 kJ.mol$^{-1}$.

### 3.2 Industrial applications of current significance

Considering first the importance of hydrodesulfurization and hydrogenation for the industrial refining of fossil fuels, we revisit here these two key reactions which have been the subject of our previous investigations[24] Figure 4a) first presents a volcano plot for hydrodesulfurization of thiophene: ordinates of data points are experimental activities for highly dispersed unsupported transition metal sulfides as reported by Lacroix et al. [25,26]. Abscissae of data points are metal-sulfur bond energies $E_{MS}$, and the optimum $E_{MS,opt}$ is found at 137.3±3 kJ.mol$^{-1}$. In ESI section 4, we also illustrate in details the relevant case of biphenyl hydrogenation[27], where the volcano plot exhibit a maximum for $E_{MS,opt}$ = 122.2 kJ.mol$^{-1}$.

Figure 4b) demonstrates that another nice volcano plot is obtained for ammonia synthesis catalyzed by carbon supported metals[28,29], using the $E_{MN}$ descriptor and revealing an optimal value of 95 kJ.mol$^{-1}$ (see ESI section 5). The same descriptor is also valid for the ammonia decomposition[30] (ESI section 5, Figure S.5.1). Other volcano plots are obtained for key reactions involved in petrochemistry, such as hydrogenation of alkenes[3,6], aromatics[3], and CO[31,32]. These cases reveal that the metal-carbon bond energy $E_{MC}$ calculated in carbide materials is the most relevant descriptor (see ESI sections 6, 7 and 8 respectively) which exhibits optimal values depending on the considered reaction varying from 69.0 kJ.mol$^{-1}$ to 117 kJ.mol$^{-1}$. Last but not least, a volcano plot was recently reported for selective oxidation of methane catalyzed by orthophosphates and pyrophosphates of transition metals[33]. These catalysts are described again by $E_{MO}$ , with $E_{MO,opt}$ = 160 kJ.mol-1.

In summary, for the twelve target reactions considered here, we have identified five main descriptors $E_{MO}$, $E_{MH}$, $E_{MS}$, $E_{MN}$ and $E_{MC}$ of which optimal values $E_{MX,opt}$ found for the best catalysts span a rather wide energy interval from 33 kJ.mol$^{-1}$ to 178 kJ.mol$^{-1}$. Due to the large chemical diversity within these reactions, this wide range is not so surprising. Nevertheless, it would be of first relevance if we could identify a rationale for this variation of optimal M-X bond energy as a function of the targeted reaction. It is the goal of the next sections to address this challenging question.

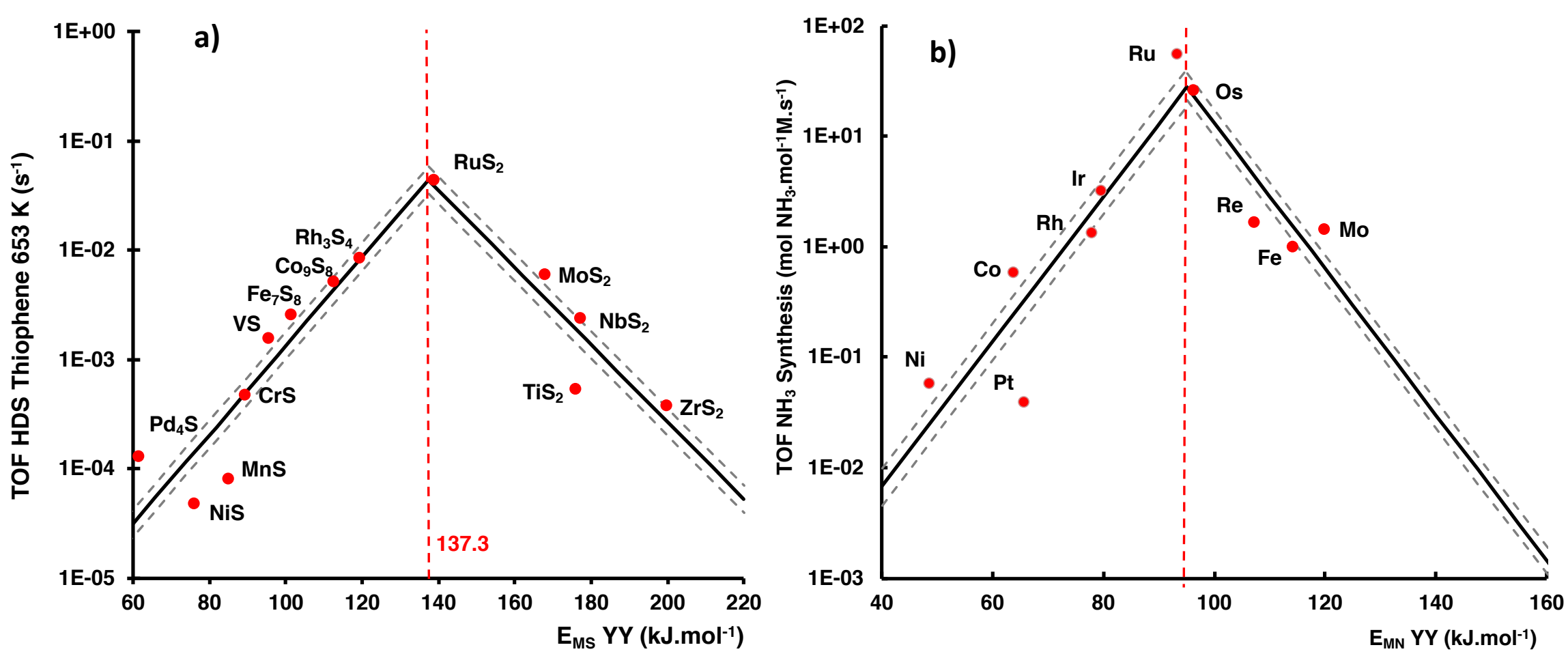

**Figure 4** a) Volcano plot for thiophene hydrodesulfurization. Experimental activity patterns in ordinates from [19] and [20]. b) volcano plot for the synthesis of ammonia by carbon supported metals (see section 4 in Supplementary Materials). In ordinates are reported (log scale) the experimental data from Ozaki [28], also used by Jacobsen et al. [23,2] . In abscissae, $E_{MS}$ and $E_{MN}$ values computed according to the Yin Yang method for bulk unit-cells of the various sulfides and nitrides indicated. In both cases, dotted lines bracket the left- and right-hand side regression lines by ± the standard deviation of experimental ordinates with respect to their projection on the corresponding regression line.

### 3.3 Comments on the choice of $E_{MX}$ descriptors

In accordance with the principle of Sabatier, the proper description of a pattern of catalytic activities exhibited by a consistent set of transition metals compounds should involve quantitative measures of interactions strengths between surface sites and adsorbed reactants, intermediates and products. Along the reaction pathways, some elementary steps will critically determine the overall rate, through the magnitude of the activation barrier for the corresponding bond dissociation or bond formation. Active sites at the surface of transition metal based heterogeneous catalysts are commonly viewed as undercoordinated metal ions $M^{\delta+}$, so that a fruitful analogy with the teachings of organometallic chemistry can help representing the activation processes, where adsorbed species are considered as transient ligands. Choosing the proper $E_{MX}$ descriptor is then equivalent to identifying along the overall pathway the molecular ligand bearing the critical bond, insufficiently activated if this ligand interacts too weakly with the M center, while the turnover frequency sharply decreases as the interaction strength increases beyond an optimal value, tending to block the active site. $E_{MX}$ should therefore be representative of the proper ligand-M center interaction.

In the cases when operating condition stabilize well-defined transition metal compounds, like particular oxides, sulphides or nitrides, although the operando surfaces might differ somewhat from the well crystallized bulk , the choice of our Yin-Yang descriptors $E_{MO}$ , $E_{MS}$ and $E_{MN}$ calculated on the corresponding bulk crystals is rather natural, as shown rather fairly successful in sections 3.1 and 3.2.

In the cases of hydrogenation reactions of unsaturated hydrocarbons (as benzene, biphenyl, styrene, ethylene) reported in the present work or in [3] catalysed by the transition metals, one should consider the whole reaction pathway for hydrogenation: adsorption, monohydrogenation steps, and desorption. If we first consider the transition state (TS) for each monohydrogenation step, it is a three-centers complex involving one metal-hydrogen (MH), one metal-carbon (MC), and one carbon-carbon (CC) bond. For a given hydrocarbon molecule (constant CC bond energy), the fact that the MC bond energy is the predominant descriptor correlating with catalytic activity can be justified by the fact that MC is the key chemical bond which control the stability of the three-centers TS complex in order to promote the hydrogen atom addition into the CC bond. If the MC bond is too strong, the H atom addition will be prevented. If the MC bond is too weak, the CC bond cannot be activated. In addition, considering also the adsorption and desorption steps, the MC bond energy is a good descriptor correlating with the adsorption energy of hydrocarbons (as shown in [3]) which also explains why the $E_{MC}$ descriptor is key for these steps. With that respect, if $E_{MC}$ is too strong, the desorption of a hydrocarbon molecule may be rate limiting. If it is too weak, the adsorption is limiting. So, an intermediate $E_{MC}$ is expected to be optimal. By contrast, $E_{MH}$ contributes to the adsorption/desorption steps but it impacts far less the energetic change of these steps (at least for a given $pH_2$ condition). As shown in [3], $E_{MC} \sim 2E_{MH}$ as computed for carbides and hydrides in the NaCl structure respectively.

Similarly, in the case of CO methanation catalysed by transition metals, the critical bond to dissociate is C≡O, while organometallic chemistry teaches that metal carbonyls are systematically bonded through the carbon atom, justifying again the choice of $E_{MC}$ as descriptor rather than $E_{MH}$ or $E_{MO}$.

For benzene and CO hydrogenation reactions catalysed by metals discussed above, we present in ESI section 10 examples of activity patterns plotted against alternatives to $E_{MC}$ which fail completely to recover the volcanoes well described by the latter.

When catalytic $M^k_{i_k}X_j$ materials including more than one transition metal $M^k$ are considered, or if metal ions are found at inequivalent positions in the structure, there are several possible target metals in the Yin-Yang calculation. Since in silico screening for innovative catalysts will preferably address alloys or complex multi-elemental compositions, instead of the pure metals or their binary combinations with non-metals, it is important to figure out guidelines for identifying the relevant targets. One must be aware that establishing the experimental volcano for a given reaction with a training set of monometallic catalysts is a crucial step for practical applications of any descriptor-based discovery strategy, unless another rule or principle allows to describe a priori the optimal catalysts, as proposed in the present report. Since the calculation of Yin-Yang descriptors is straightforward and fast, the strategy we can recommend is to compute all possible $E_{M^kX}$ values, and provided the volcano curve can be a priori outlined by a set of monometallic catalysts, look for the metal or crystallographic position for which the experimental catalytic activity is best positioned on the volcano by the corresponding $E_{M^kX}$ . For instance, as illustrated in figure 1a) in the section 3.1, the catalytic activity of the perovskite $La_2MnNiO_6$ in OER is well described by $E_{MnO} = 182.7\ kJ.mol^{-1}$, situating this material on the right hand side slope of the volcano. However, for this perovskite, as shown on figure 2 and in Table ST.1.1 of ESI, $E_{NiO} = 132.4\ \ kJ.mol^{-1}$, which would rather situate the material on the left hand side slope without significantly displacing the crossing point localizing the optimum. The comparable monometallic perovskites $LaMnO_3$ and $LaNiO_3$ have respectively $E_{MnO} = 175.8\ kJ.mol^{-1}$ and $E_{NiO} = 147.3\ \ kJ.mol^{-1}$, indicating a weak mutual influence of Mn and Ni ions binding in the bimetallic perovskite, and that the corresponding surface sites may be considered to exhibit relatively independent catalytic activities. Alternatively, a composition weighted average (lever rule) in the set of $E_{M^kX}$ may provide the correct description, as illustrated for binary alloys $Ir_3Cu$, $Cu_3Ir$ and $Re_3Ir$ in benzene hydrogenation in Figure S.7.1 of ESI. It appears to be also the case for the perovskite $La_2CrNiO_6$ (Table ST.2.2 and Figure S.2.2 in ESI), in contrast to $La_2NiMnO_6$ discussed above. We have also discussed this issue in section 1.1.1 of our book on catalysis by transition metal sulphides[24], a domain for which the lever rule appears also nicely effective.

In summary, our "Yin-Yang" $E_{MX}$ descriptors should be considered as falling in the broad category of approximations of the exact interaction energies. However, we will show in section 3.5 that they can moreover provide estimates of the catalyst's average surface energies operando and as such allow identifying the optimal catalysts.

### 3.4 Illustration of a general relationship between optimal $E_{MX}$ and enthalpies of reaction

Let us consider the standard enthalpy of the reaction $R \rightarrow P$ and introduce the quantity $\Delta H^{0,X}{}_{RP}$ defined by:

$$\Delta H^{0}{}_{RP} = \ tn_{RX_s} \Delta H^{0,X}{}_{RP} \tag{2}$$

where $n_{RX_s}$is the stoichiometric number of a reactant of reference $RX_s$ containing the describing element X, with stoichiometric indice $s$. We define$\Delta H^{0,X}{}_{RP}$ as the *minimal* value of the standard enthalpy of reaction $R \rightarrow P$ , per mole of X atoms transferred from $R$ to $P$, where $R$ means the sum of reactants including $RX_s$ , and $P$ means the sum of products. We distinguish reactions leading to a dissociation of $RX_s$ into $s$ moieties among products, from those leading to a product $R'X_s$ i.e. not dissociative. In the first case we set $t = s$ , and $t = 1$ in the second. With these conventions it is always possible to normalize standard enthalpies of reaction $\Delta H^{0}{}_{RP}$ by setting $tn_{RX_s} = 1$ as shown with the following examples:

Ammonia synthesis and decomposition form a couple of direct and inverse reactions implying respectively the dissociative chemisorption and associative desorption of $N_2$. Choosing N as the describing element, they may be written so that $sn_{RX_s} = 1$ as:

Ammonia synthesis: ½ $N_2$(g) + 3/2 $H_2$(g) = $NH_3$(g)

$RX_s = N_2$ therefore $n_{RX_s} = 1/2$ and $s = 2$

Ammonia decomposition: $NH_3$(g) = ½ $N_2$(g) + 3/2 $H_2$(g)

$RX_s = NH_3$ therefore $n_{RX_s} = 1$ and $s = 1$

For these two reactions, $|\Delta H^{0,N}{}_{RP}|$ is the same.

Another couple of direct and inverse reaction is formed by oxygen evolution and oxygen reduction, implying respectively associative desorption and dissociative chemisorption of $O_2$' Choosing O as the describing element, the following equations have same $|\Delta H^{0,O}{}_{RP}|$ and $sn_{RX_s} = 1$:

Oxygen evolution: OH(-a) + H(+a) = 1/2$O_2$(g) + $H_2$(g)

$RX_s = OH(-a)$ therefore $n_{RX_s} = 1$ and $s = 1$

Oxygen reduction: $1/2O_2$(g) + $H_2$(g) = OH(-a) + H(+a)

$RX_s = O_2$ therefore $n_{RX_s} = 1/2$ and $s = 2$

By contrast, the reactions of unsaturated hydrocarbons (benzene, ethene, styrene) hydrogenations do not imply C-C bonds dissociations, so that choosing C as the describing element they can be written with $n_{RX_s} = n_{H_tC_s} = 1$ but while $s = 6, 2,$ $or\ 8$, we retain $t = 1$:

Hydrogenation of benzene into cyclohexane:

$C_6H_6$(g) + $3H_2$(g) = $C_6H_{12}$(g)

Hydrogenation of ethene into ethane:

$C_2H_4$ (g) + $H_2$(g) = $C_2H_6$(g)

Hydrogenation of styrene into ethyl benzene:

$C_8H_8$ + $H_2$(g) = $C_8H_{10}$

In what follows we provide the experimental standard heats of reaction $\Delta H^{0,X}{}_R$ for the 12 reactions considered, extracted from the database provided with the software HSC chemistry V7.1[34]. In the following reaction equations, (g) stands for gas phase, (-a) stands for aqueous anion, (+a) stands for aqueous cation. The reactant of reference is underlined.

*OER and Photocatalytic OER:*

<u>OH(-a)</u> + H(+a) = $1/2O_2$(g) + $H_2$(g)

$\Delta H^{0,O}{}_{RP} = +354.6\ kJ.mol^{-1}$ X=O, $n_{RX_s} = 1$ and $s = 1$

*ORR (the reverse reaction of OER)*

$1/2\underline{O_2}$(g) + $H_2$(g) = OH(-a) + H(+a)

$\Delta H^{0,O}{}_{RP} - 354.6\ \ kJ.mol^{-1}$ X=O, $n_{RX_s} = 1/2$ and $s = 2$

*Selective oxidation of methane:*

$\underline{CH_4(g)} + O_2(g) = CO_2(g) + 2H_2(g)$

$\Delta H^{0,C}{}_{RP} - 318.9\ \ kJ.mol^{-1}$ X=C, $n_{RX_s} = 1$ and $s = 1$

*Hydrodesulfuration of thiophene:*

$\underline{C_4H_4S(g)} + 4H_2(g) = C_4H_{10}(g) + H_2S(g)$

$\Delta H^{0,S}{}_{RP} = -286.4\ kJ.mol^{-1}$ X=S, $n_{RX_s} = 1$ and $s = 1$

*Hydrogenation of biphenyl under $H_2$+$H_2S$ atmosphere:*

$C_{12}H_{10}(g) + H_2S(g) = \underline{C_{12}H_8S(g)} + 2H_2(g)$

$\Delta H^0{}_R = -41.4\ kJ.mol^{-1}$

(Formation of dibenzothiophene)

$\underline{C_{12}H_8S(g)} + 5H_2(g) = C_{12}H_{16}(g) + H_2S(g)$

$\Delta H^0{}_R - 180.1\ kJ.mol^{-1}$

(HDS of dibenzothiophene to phenylcyclohexane)

$\Delta H^{0,S}{}_{RP} = -221.5\ kJ.mol^{-1}$ X=S, $n_{RX_s} = 1$ and $s = 1$

Both reactions are thermodynamically favoured in the reaction conditions (in presence of $H_2S$ to stabilize TMS catalyst) and should occur serially. Therefore we retain the sum of the two standard heats of reaction.

*Methanation of CO:*

$\underline{CO(g)} + 3H_2(g) = CH_4(g) + H_2O(g)$

$\Delta H^{0,C}{}_{RP} = -205.9\ kJ.mol^{-1}$ X=C, $n_{RX_s} = 1$ and $s = 1$

*Synthesis of ammonia:*

$1/2\underline{N_2(g)} + 3/2H_2(g) = NH_3(g)$

$\Delta H^{0,N}{}_{RP} = -46.0\ kJ.mol^{-1}$ X=N, $n_{RX_s} = 1/2$ and $s = 2$

*Decomposition of ammonia:*

$\underline{NH_3}(g) = 1/2N_2(g) + 3/2H_2(g)$

$\Delta H^{0,N}{}_{RP} = 46.0\ kJ.mol^{-1}$ X=N, $n_{RX_s} = 1$ and $s = 1$

*Hydrogenation of benzene into cyclohexane:*

$\underline{C_6H_6(g)}$ + $3H_2(g) = C_6H_{12}(g)$

$\Delta H^{0,C}{}_{RP} = -206\ kJ.mol^{-1}$ X=C, $n_{RX_s} = 1$ and $s = 1$

*Hydrogenation of ethene into ethane:*

$\underline{C_2H_4\ (g)}$ + $H_2(g) = C_2H_6(g)$

$\Delta H^{0,C}{}_{RP} = -136.4\ kJ.mol^{-1}$ X=C, $n_{RX_s} = 1$ and $s = 1$

*Hydrogenation of styrene into ethybenzene*

$\underline{C_8H_8(STYl)}$ + $H_2(g) = C_8H_{10}(EBZl)$ $\Delta H^0{}_R = -116.3\ kJ.mol^{-1}$

$\underline{C_8H_8(STYg)}$ + $H_2(g) = C_8H_{10}(EBZg)$ $\Delta H^0{}_R = -118.5\ kJ.mol^{-1}$

$\underline{C_8H_8(STYg)}$ + $H_2(g) = C_8H_{10}(EBZl)$ $\Delta H^0{}_R = -160.9\ kJ.mol^{-1}$

Standard enthalpies of this reaction depend whether the aromatics are transferred or not from gas to liquid phase. A flash calculation based on the Peng-Robinson equation at 315 K and 35 bar $H_2$ pressure (the testing conditions reported by Corvaisier et al.[7]) reveals that about 50% of ethylbenzene is in liquid phase at thermodynamic equilibrium. Since the silica support used in Corvaisier et al. experiments has a specific area of 300 $m^2.g^{-1}$ and pore volume 1.15 $cm^3.g^{-1}$, the average pore diameter is about 15 nm. One expects therefore the liquid product to fill up these pores partially by capillarity. We consider therefore the average:

$\Delta H^{0,C}{}_{RP} = -131.9\ kJ.mol^{-1}$ X=C, $n_{RX_s} = 1$ and $s = 1$

*HER:*

H(ads) + $\underline{H(+a)}$ + $e^-$ = $H_2(g)$

$\Delta H^{0,H}{}_{RP} = +72.8\ kJ.mol^{-1}$ X=H, $n_{RX_s} = 1$ and $s = 1$

This follows the proposal by Trassati[22] that the HER process can be written as a two-step process:

Discharge step: $H(+a) + e^- = H(ads)$ (a)

Associative desorption: $2\ H(ads) = H_2(g)$ (b)

It was first proposed in a general form by Parsons[35] as quoted by Trassati. HER would then be described by an Eley-Rideal (or Volmer-Heyrovsky) mechanism by which an aqueous proton discharges an adsorbed H, which induces an associative desorption of dihydrogen. We have besides[34]:

$H_2(g) = 2H(+a) + 2e^-$ $\Delta H^0{}_R = -0.088\ kJ.mol^{-1}$ (c)

$H^-(g) = H°(g) + e^-$ $\Delta H^0{}_R = +72.8\ kJ.mol^{-1}$ (d)

Since in transition metal hydrides H has an anionic character, it makes sense to situate the discharge step at the interface, and assimilate process (d) and the reverse of process (a), while the associative desorption (b) would be assimilated to the reverse of the almost athermal process (c). Therefore:

$H(ads) + H(+a) + e^- = 2H(+a) + 2e^- = H_2(g)$

The overall process has enthalpy of (d)-(c), i.e. very close to (d). With the interfacial analogous to step (d) to overcome, the HER reaction is endothermic. Indeed, the hydridic chemisorbed hydrogen should be negatively charged when the cathode is brought to a negative potential w.r.t. SHE. The recombination of two proximal surface $H^-(ads)$ into $H_2$ will be prevented by coulombic repulsion unless each surface $H^-$ is discharged of its extra electron by a proximal proton in the aqueous side of the interface.

Table 1 summarizes the reactions considered, and the corresponding $|\Delta H^{0,X}{}_{RP}|$ and $E_{MX,opt}$ values used to build Figure 5.

**Table 1:** Summary of reactions, and the corresponding $|\Delta H^{0,X}{}_{R}|$ and $E_{MX,opt}$ values (in kJ.mol-1) used to build Figure 5. Target atom for Yin-Yang calculations on MiXj model catalysts. (a): average OER and POER.

| Reaction | $\lvert\Delta H^{0,X}{}_{RP}\rvert$ | $E_{MX,opt}$ | X | $n_{RX_s}$ | $s$ |
|---|---|---|---|---|---|
| OER | 354.6 | 177.6 (a) | O | 1 | 1 |
| ORR A | 354.6 | 171.7 | O | 1/2 | 2 |
| Selective oxidation of methane | 318.9 | 160.0 | O | 1 | 1 |
| HDS Thiophene | 286.4 | 137.3 | S | 1 | 1 |
| HYD Biphenyl | 221.5 | 122.2 | S | 1 | 1 |
| Methanation of CO | 205.9 | 115.3 | C | 1 | 1 |
| Synth. $NH_3$ | 46 | 95.0 | N | 1/2 | 2 |
| Dec. $NH_3$ | 46 | 88.0 | N | 1 | 1 |
| HYD Benzene | 206.0 | 110.9 | C | 1 | 1 |
| HYD Ethene | 136.4 | 69.0 | C | 1 | 1 |
| HYD Styrene in EB | 131.9 | 69.5 | C | 1 | 1 |
| HER | 72.8 | 32.9 | H | 1 | 1 |

Figure 5 displays the correlation obtained between optimal $E_{MX,opt}$ values computed according to the '"Yin Yang" method, describing the optimal catalyst for the set of twelve catalyzed reactions described above, and $|\Delta H^{0,X}{}_{RP}|$ for the same reactions.

A linear correlation is found for the four following M-X bond energy descriptors:

- $E_{MO,opt}$, for oxygen evolution and reduction reactions catalyzed by transition metal oxides and selective oxidation of methane catalyzed by orthophosphates and pyrophosphates of transition metals
- $E_{MS,opt}$ for hydrodesulfurization of thiophene and hydrogenation of biphenyl catalyzed by transition metal sulfides,
- $E_{MC,opt}$ for transition metals catalyzed hydrogenations of CO into methane, of a double bond (ethene to ethane, styrene to ethylbenzene), and of an aromatic bond (benzene into cyclohexane),
- $E_{MH,opt}$ for the electrocatalyzed hydrogen evolution reaction.

Excluding $NH_3$ synthesis and decomposition, the regression line (in black) exhibits a slope of 0.5068 ~ 1/2 with a rather good squared coefficient of correlation $R^2$=0.9748. This relationship reveals that a connection exists between $E_{MX,opt}$ and $|\Delta H^{0,X}{}_{RP}|$ such as:

$$E_{MX,opt} = |\Delta H^{0,X}{}_{RP}|/2 \quad (3)$$

$NH_3$ synthesis and decomposition are clearly outliers but a similar connection holds with $E_{MN,opt} = 2|\Delta H^{0,X}{}_{RP}|$ .

We propose in the following section a theoretical interpretation of these observations.

To summarize, in this section we started by defining a normalization of standard enthalpies of reaction allowing to refer them to one mole of the describing element X transferred from reactants to products, independently of the stoichiometric numbers. Further, we established the set of 12 normalized enthalpies corresponding to the 12 activity patterns described by EMX in sections 3.2 and 3.3. Finally we presented the empirical correlation obtained between optimal values of $E_{MX}$ and the absolute value of these normalized enthalpies. We commented the outlying but coincident data obtained for ammonia synthesis and decomposition.

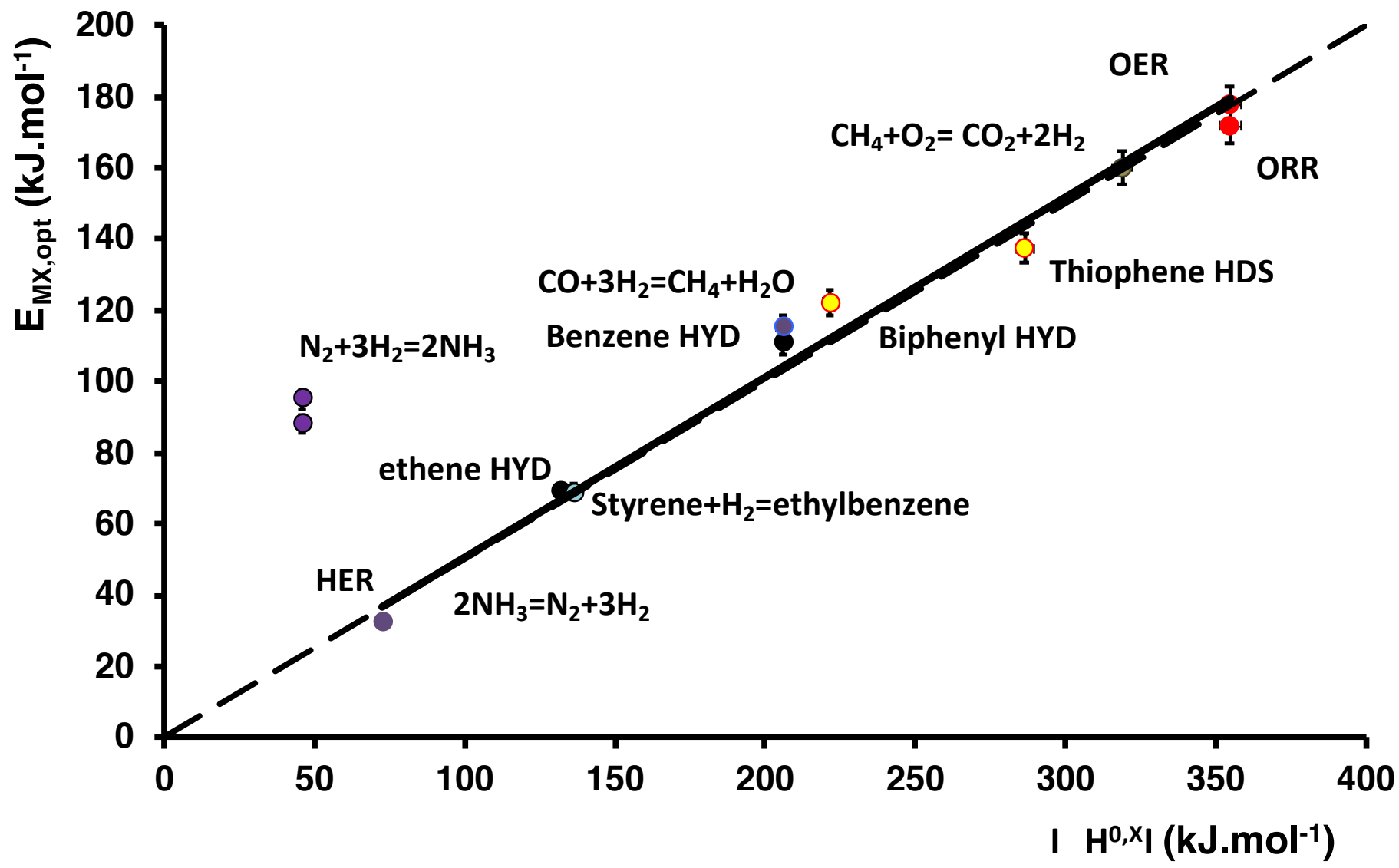

**Figure 5**: Linear relationship obtained between optimal DFT bond energies $E_{MX,opt}$ in volcano plots, as determined by the Yin Yang method (ordinates, errors bars fixed at an average of 3%), and the quantity $|\Delta H^{0,X}{}_{RP}|$ derived from absolute values of tabulated experimental standard enthalpies of the corresponding reactions. The continuous black plot is the regression line: y=0.5068x ($r^2$=0.9748), excluding outliers $NH_3$ synthesis and decomposition, while the theoretical prediction $1/2|\Delta H^{0,X}{}_{RP}|$ is the dashed line. This illustration includes direct and reverse reactions (OER and ORR, and $NH_3$ synthesis and decomposition) for which we verify the coincidence of descriptors of optimal catalysts $E_{MX,opt}$. For $NH_3$ synthesis and decomposition outliers $E_{MX,opt} \approx 2\,|\Delta H^{0,X}{}_{RP}|$

### 3.5 Theory of the optimal catalyst

Closing a Born-Haber cycle between reactants and products in fluid phase and adsorbed phase on catalyst $j$ surface, the free energy balance can be written:

$$\Delta \mathrm{G}_{RP} = \Delta G_{RP,j} + \Delta G_{R,j}^{ads} - \Delta G_{P,j}^{ads} \qquad (4)$$

where $\Delta G_{RP,j}$ is the difference in free energy between adsorbed products and reactants on catalyst $j$, $\Delta G_{R,j}^{ads}$ and $\Delta G_{P,j}^{ads}$, the summed free energies of adsorption of reactants and products respectively on catalyst $j$ surface, and $\Delta G_{RP}$ the difference in free energy between reactants and products in fluid phase.

Keeping in mind that any adsorption process can be decomposed into the sequential physisorption and chemisorption steps, we have:

$$\Delta G_{R,j}^{ads} = \Delta G_{R,j}^{ads,\varphi} + \Delta G_{R,j}^{ads,\chi} \quad (5)$$

And:

$$\Delta G_{P,j}^{ads} = \Delta G_{P,j}^{ads,\varphi} + \Delta G_{P,j}^{ads,\chi} \quad (6)$$

Then, the reaction pathway in adsorbed phase can be formally decomposed into $N$ consecutive elementary steps indiced $k, 1 \leq k \leq N$ so that:

$$\Delta G_{RP,j} = \sum_1^N \Delta G_{j,k} \quad (7)$$

This sequence includes the rate determining step (r.d.s.) of index $L$, $1 \leq L \leq N$. Moreover, we may identify steps 1 and $N$ to the chemisorption of reactants and reverse chemisorption of products respectively.

Further, as shown valid in our previous theoretical investigations[3], let us introduce scaling linear relationships between $\Delta G_R^{ads}(0K)$, $\Delta G_P^{ads}(0K)$ and $E_{MX}$, the Ying-Yang M-X bond energy for the $j$ catalyst $M_iX_j$:

$$d\Delta G_{R,j}^{ads}(0K) = -rdE_{MX,j} \quad (8)$$

$$d\Delta G_{P,j}^{ads}(0K) = -pdE_{MX,j} \quad (9)$$

with $r > 0$ and $p > 0$.

Moreover, formally, since reactants and products will not adsorb on the virtual catalyst described by $E_{MX,0} = 0$:

$$\Delta G_P^{ads}(E_{MX} = 0{,}0K) = \Delta G_R^{ads}(E_{MX} = 0{,}0K) = 0 \quad (10)$$

Plugging equations (8) and (9) into equation (4):

$$\Delta G_{RP,j}(0K) = tn_{RX_s}\Delta G_{RP}{}^{X}(0K) + (r-p)E_{MX,j} \quad (11)$$

where $\Delta_{RP}G^X(0K)$ refers according to (1) to one mole of X atoms transferred from $R$ to $P$.

Then assuming that the Brønsted-Evans-Polanyi relationship (BEP) is of general validity for any elementary step of the reaction pathway connecting the adsorbed reactants $R$ to the adsorbed products $P$, it can be expressed as:

$$\Delta G_{j,k}^{\pm} = \left(\gamma_{j,k} + a\right)\left|\Delta \mathrm{G}_{j,k}\right| + \Delta G_{0,j,k}^{\pm} \quad (12)$$

where $\Delta G_{j,k}^{\pm}$ stands for the reaction barrier for each step $k$ on catalyst $j$, $\gamma_{j,k} > 0$ for the Polanyi reaction coefficient, and $a = 0$ for $\Delta \mathrm{G}_{j,k} < 0$ , $a = 1$ for $\Delta \mathrm{G}_{j,k} > 0$ . $\Delta G_{0,j,k}^{\pm}$ is the common intercept of the linear relationships with the axis of ordinates at $\Delta \mathrm{G}_{j,k} = 0$, it also represents the intrinsic, or minimal activation barrier of the considered reaction. $\Delta \mathrm{G}_{j,k}$ is the free energy difference between adsorbed reactants and products of step $k$ on catalyst $j$. Equation (12) formulates the BEP relationship for both the direct and reverse reactions of any step $k$.

At this point, we introduce the following hypothesis: the validity of the same BEP relationship for the sequence of elementary steps in adsorbed phase for each catalyst in the family of analogs comprising an optimal catalyst, so that the coefficients in equation (12) are independent of $k$ and $j$.

In other terms, we assume that it depends merely on the particular (gas phase) reaction considered, catalyzed by $M_iX_j$ solid:

$$\gamma_{j,k} = \gamma \quad (13)$$

$$\Delta G^{\pm}_{0,j,k} = \Delta G^{\pm}_{0} \tag{14}$$

This hypothesis may appear relatively strong for the 12 reactions considered in the present work, but we notice that if several successive elementary steps occur in adsorbed phase, they are chemically similar since they involve predominantly mono(de)hydrogenation steps, and mono-oxidation steps in the case of oxidation of methane. So, this analysis may intuitively explain why our hypothesis is meaningful. Although we cannot prove the general validity of this hypothesis, we provide in ESI, section 9, an analysis of our recent DFT results[36] on the activation of $MoS_2$ Mo-edges by $H_2$, releasing $H_2S$ which gives evidence in this particular case and (non-optimal) transition metal sulfide catalyst.

Accordingly, for the reaction pathway in adsorbed phase on the optimal catalyst, the activation barrier for the r.d.s. is minimal with respect to non-optimal catalysts in the series of analogs, so that $\Delta G^{\pm}_{0}$ being the minimal barrier possible for the considered reaction in adsorbed phase:

$$\Delta G^{\pm}_{L,j} \geq \Delta G^{\pm}_{L,opt} = \Delta G^{\pm}_{0} \tag{15}$$

and therefore, according to equation (10):

$$\left|\Delta G_{L,opt}\right| = 0 \tag{16}$$

But since $\Delta G^{\pm}_{L,opt}$ is the barrier for the r.d.s, it is also the maximal barrier along the reaction pathway in adsorbed phase for the optimal catalyst, so that:

$$\text{For } k \neq L, 1 \leq k \leq N, 0 \leq \Delta G^{\pm}_{k,opt} \leq \ \Delta G^{\pm}_{L,opt} = \Delta G^{\pm}_{0} \tag{17}$$

It ensues that:

$$\text{For } 1 \leq k \leq N, \left|\Delta G_{k,opt}\right| = \left|\Delta G_{L,opt}\right| = 0 \tag{18}$$

So that from (7):

$$\Delta G_{RP,opt} = 0 \tag{19}$$

Finally, equation (4) reduces for the optimal catalyst to:

$$\Delta G_{RP} = \Delta G_{R,opt}^{ads} - \Delta G_{P,opt}^{ads} \tag{20}$$

In other terms, on the optimal catalyst, the free energy change from reactants to products is exactly compensated in adsorbed phase. Equation (20) implicitly accounts for $\Delta G_{R,opt}^{ads,\chi} = 0$ and $\Delta G_{P,opt}^{ads,\chi} = 0$ while equations (8) and (9) remain valid for the contributions of physisorption. We notice however that within the assumptions of our model, assuming the reactants physisorption or the reverse physisorption of products are activated, the combination of equations (8), (9), (12), (13) and (14) leads to:

$$\Delta G_j^{\pm,ads,\varphi} = \gamma r E_{MX,j} + \Delta G_0^{\pm} \tag{21}$$

And:

$$\Delta G_j^{\pm,des,\varphi} = (\gamma + 1) p E_{MX,j} + \Delta G_0^{\pm} \tag{22}$$

Being linear in $E_{MX,j}$ with positive coefficients, these equations have no minimum. Therefore it can be inferred that an optimal catalyst cannot result from one of these steps being the r.d.s. . Scheme 1 exemplifies the difference between reaction profiles for non-optimal and optimal catalysts.

*a) General case with rds, k=L*

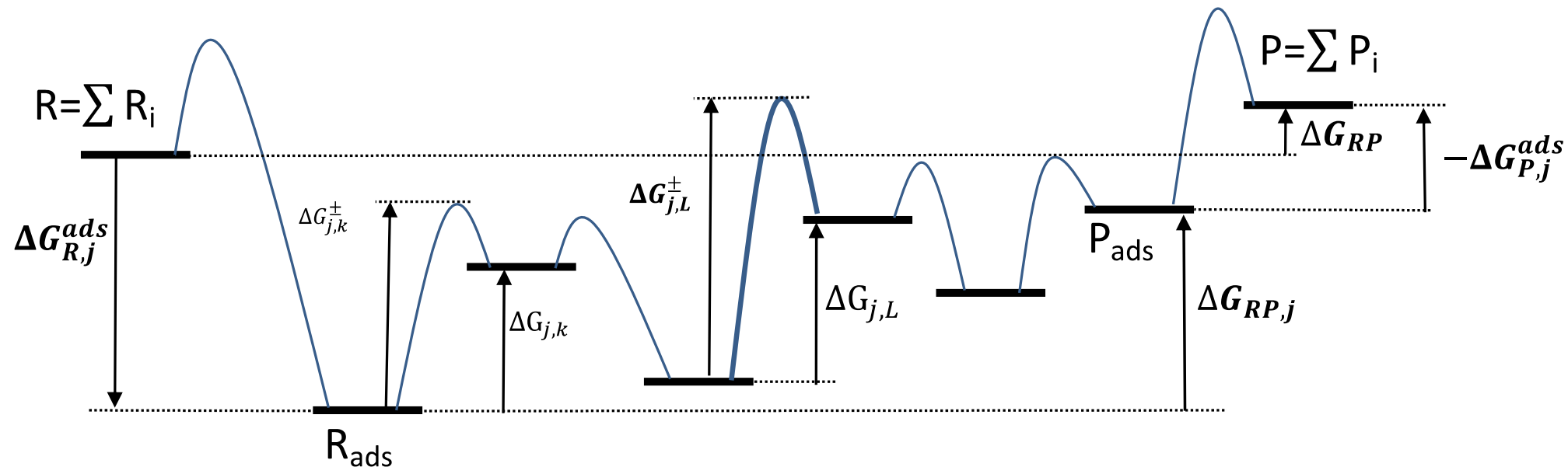

*b) Optimal catalyst with* $\Delta G_{opt,k} = 0$

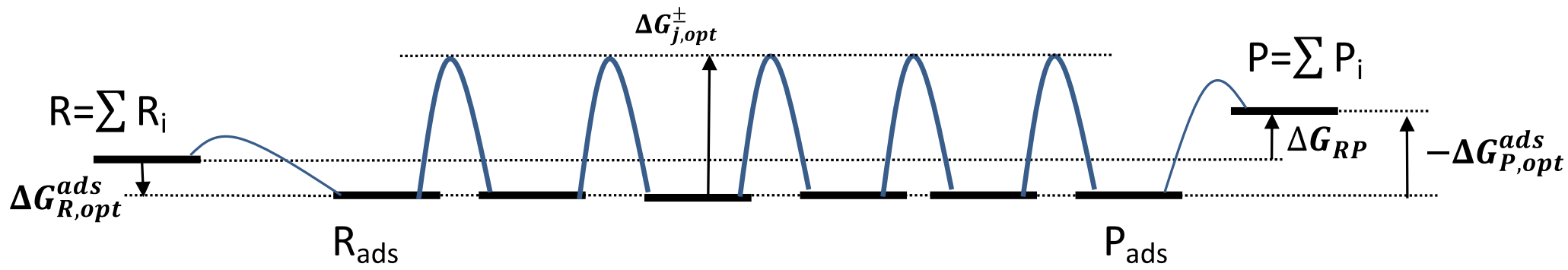

**Scheme 1 :** Energy profiles a) for the general case of a non-optimal catalyst with rate determining step k=L, and b) for the optimal catalyst for which $|\Delta G_{k,opt}| = |\Delta G_{L,opt}| = 0 \; for \; 1 \leq k \leq N$

Combining equations (11) and (19) provides:

$$tn_{RX_s} \Delta G_{RP}{}^{X}(0K) + (r-p)E_{MX,opt} = 0 \qquad (23)$$

which has a positive solution:

$$E_{MX,opt} = -\, tn_{RX_s} \Delta G_{RP}{}^{X}(0K))/(r-p) \qquad (24)$$

only and only if $\Delta G_{RP}{}^{X}(0K)$ and $(r-p)$ have opposite signs, i.e. reactants should be more stabilized than products by the catalyst, $r > p$ if $\Delta G_{RP}{}^{X}(0K) < 0$ , and products should be more stabilized than reactants by the catalyst, $r < p$ if $\Delta G_{RP}{}^{X}(0K) > 0$. Notice that these constraints apply to the catalysts, not the reaction, since R and P can be formally exchanged considering either the direct or the reverse reaction. $r = p$ would mean that this particular catalyst would not catalyze the reaction at all.

Notice also that relations (8) and (9) could be actually non-linear or approximately piecewise linear, implying for instance slopes $r_1$, $p_1$ for $E_{MX} \leq E_{MX,1}$ and $r_2$, $p_2$ for $E_{MX} >$ $E_{MX,1.}$ then equation (23) would have for instance two solutions, implying the existence of two optimal catalysts for the reaction considered. This is what we found and discussed for hydrogenation of ethene and styrene in our previous publications [3,7].

Equation (24) is close to the empirical correlation illustrated in the previous subsection section, but a priori $(r-p)$ has no reason to be a universal constant: however, it is what we will demonstrate in what follows.

First, we notice that our Yin-Yang descriptor $E_{MX,j}$ of catalyst $j$ being the cohesive energy per M-X bond, or in other terms the pairwise interaction energy between M and X, is proportional to the average surface energy $\sigma_{*,j}$, of a $M_iX_j$ catalyst particle exposing bare surfaces, and offering undercoordinated M* atoms, created by cutting (M-X) bonds crossing cleavage planes. According to the known approximation[37]

$$\sigma_{*,j} = dG_{MX,j}/dS = \lambda\langle\Delta Z\rangle E_{MX,j}/2 \qquad (25)$$

Where $\lambda$ is the areal density of M* sites in mol.m$^{-2}$, and $\langle\Delta Z\rangle = \langle Z_B\rangle - \langle Z_S\rangle$ the average difference between the average coordination number $Z_B$ of M in the bulk, and its average coordination number $\langle Z_S\rangle$ in surface. In what follows, we will assume that $\lambda$ and $\langle\Delta Z\rangle$ are the same for a family of analogous catalysts.

Since $E_{MX,j}$ is computed by DFT at 0K, $\sigma_{*,j}$ is the surface energy at 0K.

Further, let us eliminate $(r-p)$ from equation (11) thanks to (24), assuming the solution exists:

$$\left|\Delta G_{RP,j}(0K)\right| = tn_{RX_s}\left|\Delta \mathrm{G}_{RP}{}^{X}(0K)\right|\left(1 - E_{MX,j}/E_{MX,opt}\right) \qquad (26)$$

Then, introducing the continuous variable $\delta = E_{MX,j}/E_{MX,opt}$ :

$$|\Delta G_{RP}(0K,\delta)| = tn_{RX_s}\left|\Delta \mathrm{G}_{RP}{}^{X}(0K)\right|(1-\delta) \qquad (27)$$

Let us then remember that any heterogeneous catalyst must be activated “in situ” in order to function. In other terms the surface must be depassivated in order to express active sites M*, gaining therefore the surface energy $\sigma_{*,j}$. The latter becomes the driving force to adsorption of reactants R and products P.

Considering a virtual path connecting the “activated” reference catalyst of null surface energy described by $E_{MX,0} = 0$ (or $\delta = 0$) and the activated optimal catalyst described by $E_{MX,opt}$ (or $\delta = 1$), let us express the transfer of surface energy to R and P upon their adsorption on the catalyst’s surface. The free energy $dG_{RP}(0K,\delta) = \lambda\left|\Delta G_{RP,j}(0K)\right|d\delta$ represents the infinitesimal amount of surface energy transferred to reactants and products by the catalyst when its description in changed from $\delta$ to $\delta + d\delta$. It is balanced by the change of surface energy of the bare catalyst from $\delta$ to $\delta + d\delta$ when adsorbing $tn_{RX_s}$ moles of X: $\lambda tn_{RX_s}\langle\Delta Z\rangle dE_{MX,j}/2 = \lambda tn_{RX_s}\langle\Delta Z\rangle E_{MX,opt}d\delta/2$ .

The integral of $dG_{RP}(0K,\delta)$between $\delta = 0$ and $\delta = 1$ is:

$$\int_0^1 \lambda\left|\Delta G_{RP,j}(0K)\right|d\delta = \int_0^1 \lambda tn_{RX_s}\left|\Delta \mathrm{G}_{RP}{}^{X}(0K)\right|(1-\delta)d\delta$$

So that:

$$\int_0^1 \lambda\left|\Delta G_{RP,j}(0K)\right|d\delta = \lambda tn_{RX_s}\left|\Delta \mathrm{G}_{RP}{}^{X}(0K)\right|/2 \qquad (28)$$

It is balanced by the integral:

$$\int_0^1 \lambda tn_{RX_s}\langle\Delta Z\rangle E_{MX,opt}d\delta/2 = \lambda tn_{RX_s}\langle\Delta Z\rangle E_{MX,opt}/2 \qquad (29)$$

Therefore:

$$\left|\Delta G_{RP}{}^{X}(0K)\right| = \langle\Delta Z\rangle E_{MX,opt} \qquad (30)$$

Neglecting vibrational corrections to enthalpies, equation (30) translates into:

$$|\Delta H^{0,X}{}_{RP}|/\langle\Delta Z\rangle \approx E_{MX,opt} \qquad (31)$$

Equation (31) matches the relationship we have found empirically, as illustrated on Figure 5, with $\langle \Delta Z \rangle = 2$ for almost all reactions considered.

The noticeable exceptions are ammonia synthesis and decomposition, a couple of direct and inverse reactions implying respectively the dissociative chemisorption and associative desorption of $N_2$, with trivially the same $|\Delta H^{0,N}{}_{RP}|$ since we have written also the equations of these reactions so that $\mathrm{t}n_{RX_s} = 1$. We found empirically $|\Delta H^{0,N}{}_{RP}| \approx 2E_{MN,opt}$ which still matches the prediction of (31) but with $\langle \Delta Z \rangle = 0.5$ . The optimal catalysts have the same descriptors suggesting that they correspond to same transition metal nitrides. As reported in Supporting Information Tables, $Z_B = 6$ for almost all $M_iX_j$ structures considered to compute our $E_{MX}$ descriptors. We would then expect $\langle Z_S \rangle = 4$ for the catalysts of the other reactions, which is the coordination number of an ideal kink site. However, nitrides in particular are known to retain their structure down to bulk N sites occupancies as low as $\sim 0.5$ (e.g. Fm-3m $MoN_{0.506}$,[38]) corresponding therefore to $\langle Z_B \rangle \sim 3$. In such cases, $\langle \Delta Z \rangle = 0.5$ would imply $\langle Z_S \rangle \sim 2.5$ which would correspond in a non-lacunar solid to $\langle Z_S \rangle \sim 5$ , the maximal coordination number of an ideal surface site. Lacunar nitrides in particular may therefore offer less coordinated active surface sites, than are kink sites expressed by non-lacunar solids.

Notice that equation (31) implies from (24):

$$|(r - p)| = \langle \Delta Z \rangle \qquad (32)$$

So that the differential stabilizing potential of the active surface is defined by $\langle \Delta Z \rangle$. We now understand that the situation $|(r - p)| = \langle \Delta Z \rangle = 0$ would correspond to a hypothetic substance of vanishing surface energy, which could not therefore act as catalyst. Their low $\langle \Delta Z \rangle$ might be the origin of the relatively low activity of nitrides as ammonia synthesis heterogenous catalysts, while they are the stable transition metal compounds in the reaction conditions which impose high chemical potentials of dihydrogen and ammonia.

Equation (31) is found valid independently of particular reactants, products, stoichiometries and coefficients of free energy relationships. It has been derived assuming a reaction dependent, but catalyst and elementary steps independent BEP relationship, and the connection between the Yin-Yang bond energy descriptor and surface energy.

This model encompasses also situations in which chemisorption of a reactant or the reverse chemisorption of a product are identified as the rate determining step but rules out the latter being the transfer of reactants from fluid to adsorbed phase or the transfer of products from adsorbed to fluid phase.

To summarize, in this section we proposed a theoretical interpretation of the empirical correspondence presented in section 3.4. We started by giving an expression of the overall free energy balance as the sum of three parts, adsorption, transformation in adsorbed phase and desorption. Then, combining linear scaling relationships between free energies of adsorption and $E_{MX}$ and Bronsted-Evans-Polanyi linear relationships (BEP) between free energy barriers and free energy changes of elementary steps along the reaction pathway in adsorbed phase, we demonstrated that for the optimal catalyst, the free energy change between adsorbed reactants and adsorbed products is vanishing. This is a central result (Eq. 19) conditioned by the assumption that the coefficients of the BEP relationships are the same for all elementary steps along the reaction pathway. As a consequence, a simple proportionality appears between the optimal $E_{MX}$ and the normalized free enthalpy of reaction (Eq. 24). Then we showed that $E_{MX}$ is proportional to the average surface energy of the activated solid, i.e. exposing bare surface sites (Eq.25). We showed then further that since the adsorption of reactants operates a transfer of energy from the gas phase to the catalyst's surface, the proportionality in Eq. 24 becomes independent to coefficients of the scaling relationships mentioned above, but reduces to the difference of coordination numbers $\langle \Delta Z \rangle$ of the transition metal M in the bulk and in the activated surface. (Eq. 31). Finally we interpreted and commented on that basis the empirical correlation exhibited in section 3.4, i.e. $\langle \Delta Z \rangle = 2$ for 10 cases, and $\langle \Delta Z \rangle = 0.5$ for $NH_3$ synthesis and decomposition catalysed by transition metal nitrides.

## 3 Conclusion

In this report, we have shown that experimental catalytic activities for a set of twelve chemical reactions among the most important ones involved in the future production of solar fuels or in current petrochemical industry, can be organized as "volcano curves" when each catalyst is described by an intrinsic bond energy descriptor. Thereby, and in accordance with the historical Sabatier principle, an optimal catalyst can be identified for each reaction. The bond energy descriptors computed here from first principles capture the essential chemistry involved in the targeted reactions. These descriptors are also straightforward to calculate, which makes our approach a powerful tool to screen a wide range of materials and reactions.

Moreover, our work reveals a very simple direct relation between the optimal bond strength, and the absolute values of the normalized change of standard enthalpies for these key reactions. We elaborate a theoretical explanation for this observation, shown to be a consequence of chemisorbed reactants and products energy levels equalization by the optimal catalysts.

We anticipate that combining this finding with systematic DFT screening of bond strengths in bulk crystalline models of materials will be greatly helpful to discover new active catalysts, or rationalize catalysts designs, meeting immense economical and societal challenges as discussed in introduction. A refined search of catalysts close to optimality of activity but meeting also other criteria such as including if possible only non-toxic, and earth abundant elements, will be then possible, including if necessary several transition metal elements $M^k$ , and described by a particular $E_{M^k X}$, or a combination in the set, as suggested in section 3.3.

**Conflicts of interest**

There are no conflicts to declare.

**Acknowledgements**

This article is devoted to the memory of Professor Michel Che (1941-2019)

**Electronic Supplementary Information (ESI) available**

1-Correlation between EMO and occupancy of antibonding orbitals of eg parentage in the coordination sphere of transition metal cations in oxides;

2-Volcano patterns for OER and ORR reactions against EMO in transition metal oxides;

3-Volcano pattern for the hydrogen evolution reaction (HER) at transition metal cathodes against EMH in transition metal hydrides;

4-Volcano patterns for hydrodesulfurization of thiophene and hydrogenation of biphenyl against EMS in transition metal sulfides;

5-Volcano pattern for ammonia synthesis and decomposition against EMN in transition metal nitrides;

6-Volcano pattern for the selective hydrogenation of the ethyl- group in styrene, and ethene into ethane against EMC in transition metal carbides;

7-Volcano pattern for hydrogenation of benzene against EMC in transition metal carbides;

8-Volcano pattern for the methanation of carbon monoxide against EMC in transition metal carbides;

9-BEP relationship for the activation pathway of $MoS_2$ M-edge_50%S.

10-Examples of inappropriate $E_{MX}$ descriptors for hydrogenation reactions catalysed by transition metals

11-Notice on the consistency and relevance to experimental data of Yin-Yang DFT descriptors